\documentclass[aps,pra,twocolumn,showpacs,superscriptaddress]{revtex4-1}
\usepackage{amsfonts}

\usepackage{graphicx}
\usepackage{subfigure}
\usepackage{braket}
\usepackage{amsmath}
\usepackage{bm}

\begin{document}

\title{An efficient parallel method for relaxing to the minimum action wavefunction}
\author{Zachary~B.~Walters}
\affiliation{Max Planck Institute for Physics of Complex Systems, N\"othnitzer
Strasse 38, D-01187 Dresden, Germany}
\date{\today}

\begin{abstract}
  Efficient and accurate numerical propagation of the time dependent
  Schr\"odinger equation is a problem with applications across a wide range of
  physics.  This paper develops an efficient, trivially parallelizeable method
  for relaxing a trial wavefunction toward a variationally optimum propagated
  wavefunction which minimizes the propagation error relative to a platonic
  wavefunction which obeys the time dependent Schr\"odinger equation
  exactly.  This method is shown to be well suited for incorporation with
  multigrid methods, yielding rapid convergence to a minimum action solution
  even for Hamiltonians which are not positive definite.
\end{abstract}

\maketitle

As the general wave equation for nonrelativistic quantum mechanics, the time
dependent Schr\"odinger equation holds sway over a broad range of modern
physics.  However, due in part to its generality, the TDSE remains a
challenging problem to treat numerically.  For a wavefunction expanded in a
basis set, with operators represented by matrices, the wide range of physical
problems described by the TDSE makes it difficult to define propagation
procedures which are both general and efficient.  Nearly any simplifying
assumption about the form of the Hamiltonian matrix will be violated by some
problem of interest, while procedures which do not make such assumptions may
be computationally expensive.


The need for an efficient, general propagator for the TDSE has been
highlighted in recent years by the rise of strong field physics, where
problems often have particularly poor numerical properties.  A typical strong
field experiment may involve an electron tunneling free from a molecule due to
the field of an intense laser.  After reaching the continuum, the electron is
accelerated by the field of the laser and may traverse hundreds of bohr before
returning to scatter energetically from the parent ion.  The electric field
due to the laser and the continuum electron may excite the ion, while a
sufficiently energetic recollision may liberate additional electrons.  A
strong field problem may thus include singular and irregularly shaped
molecular potentials, large length scales, high energies, and time dependent,
nonperturbative external forces.  At present, calculations involving a single
electron in three dimensions reflects a significant computational challenge
\cite{schneider2006parallel}, while two electron calculations may require
thousands of processors \cite{smyth1998numerical}.

Recently, the problem of propagating a time dependent wavefunction has been
approached from the perspective of minimizing the action accumulated when the
wavefunction is expanded in a particular basis \cite{walters2010efficient}.
For a finite basis set, this action is not necessarily zero.  Minimizing this
action was shown to minimize the deviation between the propagated wavefunction
and a platonic ``true'' wavefunction which obeys the time dependent
Schr\"odinger equation exactly.  By appropriate choice of basis set, the
propagated wavefunction can be made to converge exponentially to the true
wavefunction.  The least action propagator thus represents a general method
for accurately propagating the time dependent Schr\"odinger equation.
However, the accuracy of this propagator is purchased at the cost of solving a
potentially very large linear system of equations at every time step.  For
problems involving large or unstructured basis sets, where efficient
propagators are most needed, direct solution of this linear system is likely
to prove prohibitively expensive.

This paper addresses the problem of solving the least action linear system by
developing an iterative method for decreasing the action accumulated by a
trial wavefunction.  By dividing a spacetime volume into many small
subvolumes, the accumulated action is decreased by finding corrections which
minimize the action accumulated in each subvolume.  For the common case of a
local Hamiltonian, corrections to the trial wavefunction in nonoverlapping
subvolumes are independent of one another, making this relaxation procedure
trivially parallelizeable.  Combining this relaxation procedure with multigrid
algorithms yields an overall solution procedure which replaces the solution of
a single large linear system with the solution of many small linear systems,
greatly reducing the computational cost.  As multigrid methods have
traditionally been restricted to use with linear systems arising from
elliptical PDEs, this reflects an expansion of the class of problems for which
these powerful methods are suitable.

\section{Review of the Least Action Principle}

The least action propagator was derived in \cite{walters2010efficient} as a
way to minimize the deviation between the propagated wavefunction and a
platonic ``true'' wavefunction which obeys the time dependent Schr\"odinger
equation exactly.  While \cite{walters2010efficient} employed a Taylor series
expansion for this purpose, a simpler derivation requires only the
fundamental theorem of calculus.  If the true wavefunction is given by
$\psi(x,t)$ and the approximate wavefunction is given by
$\phi(x,t)=\sum_{i,n}C_{in}\chi_{i}(\vec{x})T_{n}(t)$, the approximation error
is given by $\delta(x,t)=\psi(x,t)-\phi(x,t)$.  The norm of this deviation at
some time $t$ for a propagation step beginning at $t_{0}$ is given by
\begin{equation}
\begin{split}
\braket{\delta(t)|\delta(t)}=&\braket{\delta(t_{0})|\delta(t_{0})}
\\ &+\int_{t_{0}}^{t}dt' \frac{d}{dt'}\braket{\delta(t')|\delta(t')}.
\end{split}
\end{equation}
The assumption that $(i\frac{d}{dt}-H)\ket{\psi}=0$ can be used to eliminate
all terms involving $\bra{\psi}$ or $\ket{\psi}$ from the second term, so that
\begin{equation}
\begin{split}
  \braket{\delta(t)|\delta(t)}=&
  \braket{\delta(t_{0})|\delta(t_{0})}
  -2i\int_{t_{0}}^{t}dt'\bra{\delta(t')}H\ket{\delta(t')} \\ 
  &+2i\int_{t_{0}}^{t}dt'\bra{\phi(t')}(i\frac{d}{dt}-H)\ket{\phi(t')}.
\end{split}
\end{equation}

In this equation, terms involving $\phi(x,t)$ result from imperfectly
propagating the wavefunction, while terms involving $\delta(x,t)$ result from
the inability to represent the true wavefunction in the chosen basis.  As the
propagation error is proportional to the integral of the Lagrangian density
$L=i\frac{d}{dt}-H$, the wavefunction which minimizes propagation error is
that which minimizes the action accumulated over the time step.

The least action condition can be turned into a propagation scheme using the
calculus of variations.  For a Hamiltonian $H=H_{0}+V(x,t)$ with matrix
elements in space and time given by 
\begin{equation}
H_{i,j}=\int dx \chi_{i}^{*}(x) H_{0} \chi_{j}(x)
\end{equation}
\begin{equation}
V_{ijnm}=\int_{t}^{t+\Delta t}dt' \int dx \chi_{i}^{*}(x) T^{*}_{n}(t)
V(x,t) \chi_{j}(x) T_{m}(t)
\end{equation}
\begin{equation}
O_{i,j}=\int dx \chi_{i}^{*}(x)\chi_{j}(x)
\end{equation}
\begin{equation}
U_{n,m}=\int dt T^{*}_{n}(t)T_{m}(t)
\end{equation}
\begin{equation}
Q_{n,m}=\int dt T^{*'}_{n}(t)T_{m}(t),
\end{equation}
the action accumulated between $t$ and $t+\Delta t$ is given by
$\phi(x,t)=\sum_{in}C_{in}\chi_{i}(x)T_{n}(t)$ is given by
\begin{equation}
S = \sum_{i,j,n,m} C_{i,n}[i
  O_{i,j}Q_{n,m}-H_{i,j}U_{n,m}-V_{ijnm}]C^{*}_{j,m}
\end{equation}
and is minimized when
\begin{equation}
\frac{\partial S}{\partial C^{*}_{j,m}} = \sum_{i,n} C_{i,n}[i
  O_{i,j}Q_{n,m}-H_{i,j}U_{n,m}-V_{ijnm}]=0
\label{eq:stationaryphase}
\end{equation}
for all j,m.  Initial and boundary conditions can be specified using Lagrange
multipliers.  Defining $S'=S+\sum_{i}\lambda_{i} f^{*}_{i}$, where
$f_{i}=C_{in}T_{n}(t)-\braket{\chi_{i}|\psi(x,t)}$, minimizing $S'$ 
with the constraint that $f_{i}=0$ for all i yields the minimum action
solution with the initial condition
$\phi(x,t)=\sum_{i}\ket{\chi_{i}(x)}\braket{\chi_{i}(x)|\psi(x,t)}$. 

When using the calculus of variations in this way, it must be remembered that
it is, strictly speaking, impossible to ``minimize'' a complex quantity such
as the action.  This difficulty is particularly relevant in the context of
seeking an iterative solution procedure, as the convergence of such a
procedure can be more easily judged in the context of minimizing a real
quantity than in approaching an extremum of a functional derivative.
Accordingly, it is useful to define an action residual 
\begin{equation}
r_{jm}=-\sum_{i,n}C_{i,n}[i O_{i,j}Q_{n,m}-H_{i,j}U_{n,m}-V_{ijnm}]
\label{eq:actionresidual}
\end{equation} 
as a quantity to measure the deviation of a wavefunction from the
action-extremizing condition given by Eq. \ref{eq:stationaryphase}.  Using
this definition, the norm of the residual 
\begin{equation}
N=r_{in}O_{ij}U_{nm}r^{*}_{jm}
\label{eq:residnorm}
\end{equation} 
is
positive definite, and 
\begin{equation}
\frac{\partial N}{\partial
  C^{*}_{j,m}}=r_{in}O_{ij}U_{nm}\frac{\partial r^{*}_{jm}}{\partial
  C^{*}_{jm}}=0
\label{eq:la_residnorm}
\end{equation}
when the residual is zero and Eq. \ref{eq:stationaryphase} is
satisfied.  Restating the least action problem in this way also gives a simple
path toward improving a trial wavefunction.  If a trial wavefunction gives a
nonzero residual in Eq. \ref{eq:stationaryphase}, the expansion coefficients
must be adjusted so as to decrease the overall norm of the action residual.
Despite their technical inaccuracy, the terms ``least action'' and
``minimizing the action'' will be retained, as the terminology is familiar
from classical physics.

\section{Iterative relaxation to the least action solution}
\label{section:relaxation}

The least action propagator purchases its rapid convergence to the true
wavefunction at a high computational cost.  For a basis with $N_{x}$ spatial
and $N_{t}$ temporal basis functions, the least action propagator must solve
for $N_{x}(N_{t}+1)$ coefficients (including Lagrange multipliers) at every
time step.  For the large and irregular spatial bases where efficient
propagation schemes are most needed, the cost of directly solving the least
action linear system defined by Eq. \ref{eq:stationaryphase} is likely to
prove unacceptably high.


As an alternative to direct solution, many iterative algorithms have been
developed for the solution of linear systems of equations.  Methods such as
the conjugate gradient or generalized minimum residual algorithm attempt to
solve a general linear system $Ax=b$ by solving a smaller linear system within
a Krylov subspace\cite{greenbaum1997iterative}.

A common theme in such iterative methods is the need for a preconditioner
$M^{-1}$ which has the effect of approximately inverting the matrix $A$.
Given a trial solution $\bar{x}$ with residual $r=b-A\bar{x}$, a correction
term $\delta x=M^{-1}r$ can be used to improve the trial solution, so that a
new trial solution is given by $\bar{x}'=\bar{x}+\delta x$.  In the
limit that $M^{-1}=A^{-1}$, the correction yields an exact solution; however,
it is more typical to choose a preconditioner which is simple to calculate and
yields acceptable performance.  Generic preconditioners include Jacobi or
Gauss Seidel relaxation, while physical knowledge may allow
construction of preconditioners well suited to individual problems.

To define a preconditioner for the least action propagator, it is necessary
to return to the action integral over a volume $\mathbb{V}$
\begin{equation}
S=\int dt \int d\mathbb{V} \phi^{*}L\phi,
\end{equation}
where $L=i\frac{d}{dt}-H$ is the Lagrangian operator and the
Hamiltonian $H=T+V$ is the sum of a kinetic energy operator
$T=\frac{-1}{2m}\nabla^{2}$ and a (possibly time dependent) potential energy
operator $V$.  
The divergence theorem can then be used to express the kinetic energy operator
in terms of singly differentiable quantities
\begin{equation}
\begin{split}
\frac{-1}{2m}\int dt \int_{\mathbb{V}}d\mathbb{V} \phi^{*}\nabla^{2}\phi
=&\frac{-1}{2m}\int dt \int_{\partial \mathbb{V}} \phi^{*}\vec{\nabla}\phi\cdot d\vec{\mathbb{A}}
+ \\
&\frac{1}{2m}\int dt \int_{\mathbb{V}}d\mathbb{V} \vec{\nabla}
\phi^{*}\cdot\vec{\nabla}\phi\cdot d\vec{\mathbb{A}},
\end{split}
\end{equation}
allowing consideration of trial wavefunctions with discontinuous derivatives.
Given a Lagrangian which is a local function of the spatial coordinates and
some subvolume $\mathbb{U}$ of the original volume, the action integral can be
broken up into the action accumulated inside $\mathbb{U}$ and that accumulated
outside:
\begin{equation}
S=S_{\mathbb{U}}+S_{\bar{\mathbb{U}}},
\end{equation}
where
\begin{equation}
\begin{split}
S_{\mathbb{U}}=&\frac{-1}{2m}\int dt \int_{\partial \mathbb{U}} \phi^{*}\vec{\nabla}\phi\cdot d\vec{\mathbb{A}}
+\\ &\frac{1}{2m}\int dt \int_{\mathbb{U}}d\mathbb{U} \vec{\nabla} \phi^{*}\cdot\vec{\nabla}\phi\cdot d\vec{\mathbb{A}}+\\ 
&\int dt \int_{\mathbb{U}} d\mathbb{U} \phi^{*}V(x,t)\phi
\end{split}
\end{equation}
and $\bar{\mathbb{U}}$ is the subvolume of $\mathbb{V}$ not contained in
$\mathbb{U}$. 



It can now be seen that the requirement that the propagated function $\phi$
minimize the action accumulated over all space places restrictions on
$\phi_{\mathbb{U}}$, the function restricted to  $\mathbb{U}$.
In order for $\phi$ to minimize the action over all space,
$\phi_{\mathbb{U}}$ must minimize the action accumulated over $\mathbb{U}$,
subject to the requirement that $\phi_{\mathbb{U}}|_{\partial
  \mathbb{U}}=\phi_{\bar{\mathbb{U}}}|_{\partial \mathbb{U}}$.  If this were
not the case, it would be possible to construct a relaxed wavefunction which
would come closer to minimizing the action.

Here it is convenient to consider the norm of the least action residual.
Being positive definite, this norm allows two candidate wavefunctions to be
compared by the degree to which they minimize the action.  Let $N(\phi)$ be
the norm of the total action residual, with
$N_{\mathbb{U}}(\phi_{\mathbb{U}})$ and
$N_{\bar{\mathbb{U}}}(\phi_{\bar{\mathbb{U}}})$ the norms of the action
residuals inside and outside the volume of interest.  If there exists some
function $\varphi_{\mathbb{U}}$ for which $\varphi_{\mathbb{U}}|_{\partial
  \mathbb{U}}=\phi_{\bar{\mathbb{U}}}|_{\partial \mathbb{U}}$ and
$N_{\mathbb{U}}(\varphi_{\mathbb{U}})< N_{\mathbb{U}}(\phi_{\mathbb{U}})$,
then it is possible to construct a new global wavefunction $\phi'$, where
$\phi'_{\mathbb{U}}=\varphi_{\mathbb{U}}$ and
$\phi'_{\bar{\mathbb{U}}}=\phi_{\bar{\mathbb{U}}}$, such that
\begin{equation}
N(\phi')=N_{\mathbb{U}}(\varphi_{\mathbb{U}})+ 
N_{\bar{\mathbb{U}}}(\phi_{\bar{\mathbb{U}}})<
N(\phi)=N_{\mathbb{U}}(\phi_{\mathbb{U}})+ 
N_{\bar{\mathbb{U}}}(\phi_{\bar{\mathbb{U}}}).
\end{equation}

This logic leads naturally to a relaxation
procedure.  Given a trial solution $\bar{\phi}$ and a collection of (possibly
overlapping) subvolumes ${\mathbb{U}}$, the trial wavefunction may be
systematically improved by minimizing the action accumulated in each
subvolume, subject to the restriction that each relaxation step leave the
wavefunction unchanged on the boundary and the exterior of the relaxation
volume.  For a local Lagrangian, relaxation over two nonoverlapping volumes is
independent, allowing for parallel execution.

To cast this procedure in the form of a preconditioner, a relaxation step must
find a correction $\delta \phi_{\mathbb{U}}$ such that $L_{\mathbb{U}}\delta
\phi_{\mathbb{U}}=-L_{\mathbb{U}}\bar{\phi}$, with initial conditions $\delta
\phi(x,t_{0})=0$ and boundary conditions $\delta \phi|_{\partial \mathbb{U}}=0$.


Here it must be noted that the use of Lagrange multipliers means that the
relaxed solution $\bar{\phi}+\delta\phi$ is not guaranteed to be an extremum
of the action, but rather a critical point.  For basis functions which are
nonzero at the initial time or on the boundaries of the relaxation volume,
$\frac{\partial{S}}{\partial C^{*}_{jm}} \ne 0$, due to the nonzero Lagrange
multipliers associated with imposing the boundary conditions.  For this
reason, the relaxation procedure described here is not guaranteed to decrease
the action accumulated in the relaxation volume.  As will be seen, increasing
the size of the basis set within the relaxation volume helps to minimize this
effect.

\section{Multigrid Relaxation}
The relaxation procedure introduced in the previous section has the side
effect of introducing a length scale into the problem which does not originate
with the PDE being solved.  Because the relaxation procedure cannot change the
wavefunction outside of the relaxation volume, the size of the relaxation
volume affects the rate of convergence.  If the space is divided up into a set
of relaxation volumes of size $\mathbb{V}$, the relaxation procedure will
quickly eliminate ``rough'' error terms which accumulate action over volumes
smaller than $\mathbb{V}$, while error terms which accumulate action over
volumes larger than $\mathbb{V}$ will be eliminated more slowly.  This problem
can be rectified somewhat by relaxing over larger volumes, at the cost of
increasing the size of the linear system which must be solved.

Multigrid approaches \cite{briggs2000multigrid,trottenberg2001multigrid} seek
to eliminate this dependence on the relaxation volume by defining a treelike
hierarchical basis set.  At the base level, a single element is defined over
the entire volume of interest.  Within this element, functions are defined in
terms of some small basis set -- here, low order polynomials.  To describe
details of a function which cannot be described by this small basis set, the
element is subdivided into $n$ smaller elements, each with their own
associated basis.  Thus, the volume is divided into one element at the zeroth
level, $n$ elements at the first level, $n^{2}$ at the second level, and so
on.  This repeated subdivision produces a treelike basis, with the parents of
a volume element $e$ consisting of those elements which contain $e$ as a
subvolume and the children consisting of those elements which are a subvolume
of $e$.  The leaves of the tree consist of those elements which have not been
subdivided -- ie, the finest level of the mesh.

In defining this treelike basis set, there is no a priori need to subdivide
all volumes equally, and it may be convenient to give a compressed
representation of some function by using a few coarse elements to describe the
wavefunction in some region where it contains little detail, but many fine
elements in some region where it oscillates more rapidly.  In this way, the
multigrid basis set is closely related to wavelet decomposition, and offers
similar possibilities for reduction of memory and storage requirements.

The removal of the relaxation volume from the rate of convergence of the
multigrid relaxation procedure is accomplished by transferring the problem
between different levels of the tree, so that relaxation
may occur over many different length scales.  If $P^{N-1}_{N}$ is a
projection operator mapping a function defined on the $N$th level to a
function defined on the (coarser) $N-1$st level, and $I^{N}_{N-1}$ is an
interpolation operator transferring functions the other way, so that
$P^{N-1}_{N}I^{N}_{N-1}$ is the identity on the $N-1$st level, and
$I^{N}_{N-1}P^{N-1}_{N}$ has eigenvalues of 1 for functions which can
be expanded in the $N-1$st level basis and 0 otherwise, a linear equation
defined on one level can be projected onto another level.  For a linear system 
\begin{equation}
A^{N}_{N}x^{N}=b^{N}
\end{equation}
defined on the fine level, the projected problem on the coarse level is given
by 
\begin{equation}
  (P^{N-1}_{N}A^{N}_{N}P^{TN-1}_{N}) (P^{N-1}_{N}x^{N})=
  P^{N-1}_{N}b^{N},
\end{equation}
where $P^{TN-1}_{N}=I^{N}_{N-1}$.

Given a trial solution $\bar{x}^{N}$ with residual
$r^{N}=b^{N}-A^{N}_{N}x^{N}$, a correction $\delta x^{N}$ can be found
either by solving 
\begin{equation}
A^{N}_{N}\delta x^{N}=r^{N}
\end{equation}
on the fine grid or
\begin{equation}
  (P^{N-1}_{N}A^{N}_{N}P^{TN-1}_{N}) (P^{N-1}_{N}\delta x^{N})=
  P^{N-1}_{N}r^{N}
\label{eq:residual_rough}
\end{equation}
on the coarse grid and interpolating to the fine grid, so that
\begin{equation}
\delta x^{N}\approx I^{N}_{N-1}(P^{N-1}_{N}A^{N}_{N}P^{TN-1}_{N})^{-1}
P^{N-1}_{N}r^{N}
\label{eq:correction_fine} 
\end{equation}

The advantages of finding a correction on the coarse grid are twofold.  First,
there are fewer basis functions defined on the coarse grid, so that the
overall problem is smaller.  Second, the elements defined on the coarse grid
are larger than those defined on the fine grid by a factor of $n$, so that a
relaxation scheme such as the one described in Section
\ref{section:relaxation} will eliminate error terms which accumulate action in
volume $n\mathbb{V}$ rather than $\mathbb{V}$.  By recursively transferring
the problem between different levels of the hierarchy, error terms acquiring
action over any length scale can be made to converge, ideally yielding an
overall procedure which is rapidly convergent with rates of convergence which
do not depend on the size of the finest element.

To date, multigrid approaches have primarily been used for problems arising
from elliptical PDEs, due primarily to the lack of a suitable relaxation
method to use for corrections at each level of the hierarchy.  Relaxation
methods such as Gauss-Seidel require a Hamiltonian which is either positive or
negative definite\cite{grinstein1983multigrid}.  If the eigenvalue spectrum
spans 0, some error terms will diverge rather than converging.  For elliptical
problems, convergence is guaranteed for Gauss Seidel relaxation by the
positive definite Hamiltonian, while non elliptical problems such as the
Schr\"odinger equation require that the error terms be eliminated in some
other way, such as direct solution of the linear system at some level or
developing an improved relaxation procedure.  Multigrid approaches have been
used to solve the Schr\"odinger equation in \cite{costiner1995adaptive,
  grinstein1983multigrid, costiner1995simultaneous, chang1990multigrid}.
General reviews of preconditioning methods are given by
\cite{benzi2002preconditioning, saad2003iterative}, while
\cite{erlangga2008advances, reps2010indefinite} discuss preconditioners for
closely related Helmholtz equation.

\section{Implementation with Legendre Polynomials}

The relaxation procedure described in this paper attempts to reduce the
computational cost of performing a least action step by restricting the volume
over which the action is minimized.  Implicit in this restriction is the
assumption that reducing the volume in this way will reduce the number of
basis set coefficients to be solved for -- ie, that the chosen spatial basis
consists of localized basis functions which are nonzero only in a restricted
range.  This property is true of many popular basis sets such as finite
elements or the hierarchical multigrid bases described in the previous
section but not for, e.g., Gaussian basis sets of the kind commonly
used in quantum chemistry.  For the purposes of testing the least action
relaxation procedure, a suitable basis is provided by a set of low order
Legendre polynomials.  The Legendre basis has the advantage that many of the
matrices in the least action equation are both sparse and analytically
calculable, with well behaved matrix elements.  In one dimension, the range
$[-1,1]$ of the Legendre functions can be interpreted as a single finite
element, or the interval between two grid points in the hierarchical basis.
In one spatial dimension, the overlap matrices $O$ and $U$ are given by the
orthogonality relation
\begin{equation}
\int_{-1}^{1} dy P_{n}(y) P_{m}(y)=\frac{2 \delta_{nm}}{2n+1},  
\end{equation}
while the Q matrix is given by
\begin{equation}
Q_{nm}=\int_{-1}^{1}P'_{n}(y)P_{m}(y)=2
\end{equation}
if $n>m$ and mod$(n-m,2)=1$, 0 otherwise.
Matrix elements of the kinetic energy operator are given by
\begin{equation}
\begin{split}
T_{ij}=&\frac{1}{2m}[-P_{i}(y)P'_{j}(y)|_{-1}^{1}+\int_{-1}^{1}dy
P'_{i}(y)P'_{j}(y)] \\
=&\frac{-i(i+1)+(j^{2}+j)}{2m}
\end{split}
\label{eq:Tmatelt}
\end{equation}
when $i > j$ and mod$(i-j,2)=0$, 0 otherwise.  These formulas make use of the
identities $P_{n}(1)=1$, $P'_{n}(1)=\frac{n(n+1)}{2}$ and
$P_{n}(-y)=(-1)^{n}P_{n}(y)$.
For the purposes of this paper, $V_{0}$ is assumed to be constant
throughout the volume of interest, so that matrix elements of the potential
energy are proportional to the overlap matrix.


Intergrid transfer operators are constructed by subdividing the $[-1,1]$ range
of the Legendre polynomials into two smaller elements, one ranging from
$[-1,0]$, the other from $[0,1]$, and defining Legendre bases in the two
subelements.  Because $P_{n}(\frac{y' \pm 1}{2})$ is an $n$th order polynomial
over the range of both subelements, each basis function of the large element
can be expanded as the sum of basis functions defined in the small elements,
yielding interpolation matrices
\begin{equation}
I_{ij}^{(L)}=\int_{-1}^{1} dy' P_{i}(\frac{y'-1}{2})P_{j}(y')
\end{equation}
for the small element on the left and
\begin{equation}
I_{ij}^{(R)}=\int_{-1}^{1} dy' P_{i}(\frac{y'+1}{2})P_{j}(y')
\end{equation}
for the small element on the right, which convert from the coarse to the fine
basis.  Projection matrices are simply the transpose of the interpolation
matrices, normalized so that $(P^{(L)}+P^{(R)})\cdot(I^{(L)}+I^{(R)}) =
\mathbb{I}$ is the identity matrix in the parent basis.
\begin{equation}
P^{(L/R)}_{ij}=\frac{I^{(L/R)}_{ik}O^{(L/R)}_{kj}}{I^{(L)}_{ik}O^{(L)}_{kj}I^{(L)}_{ji}+I^{(R)}_{ik}O^{(R)}_{kj}I^{(R)}_{ji}}. 
\end{equation}

In defining these matrices, the integrals of the dimensionless parameter $y$
range from $-1$ to $1$.  For an element ``box'' of size $\Delta x \Delta t$,
these integrals acquire dimension, with $T_{ij} \propto 2/\Delta x$, $O_{ij}
\propto \Delta x/2$, $Q_{nm} \propto 2$ and $U_{nm} \propto \Delta t/2$, and
the least action equation becomes
\begin{equation}
C_{in}[i O_{ij}Q_{nm}\Delta x - T_{ij}U_{nm}\frac{\Delta t}{\Delta
  x}-O_{ij}U_{nm}V \Delta x \Delta t]=0 \forall j,m.
\end{equation}
Defining $p_{\text{max}}=\frac{1}{\Delta x}$, this
equation can be recast as
\begin{equation}
C_{in}[i O_{ij}Q_{nm} - 2 \kappa T_{ij}U_{nm}-\nu O_{ij}U_{nm}]=0 \forall j,m,
\label{eq:leastaction_dimensionless}
\end{equation}
where $\kappa=\frac{p_{\text{max}}^{2}\Delta t}{2}$ is the action due to kinetic
energy acquired by a particle with momentum $p_{\text{max}}$ in time $\Delta
t$ and $\nu=V \Delta t$ is the action acquired due to potential energy.

Parameterizing the least action equation in this way helps to clarify the role
of intergrid transfer operators in speeding
convergence.  If the size of the box is doubled, as might be seen in the
transfer from a fine to a coarse grid, $\kappa \rightarrow \kappa/4$, while
$\nu$ is unaffected.  Transferring from a fine to a coarse grid is thus
isomorphic to propagating for a time $\Delta t \rightarrow \Delta t/4$ in a
potential $V \rightarrow 4V$.  As the problem is transferred from fine to
coarse grids, the kinetic energy term will disappear, while transferring from
coarse to fine grids will make this term dominate.  If faster convergence is
needed on a grid where the potential term dominates, it may be desirable to
accomplish a single timestep in two steps of size $\Delta t/2$, which will
have the effect of mapping $\nu \rightarrow \nu/2$ and $\kappa \rightarrow \kappa/2$.

Because the Legendre polynomials are defined only within the confines of a
single element box, it is necessary when working in the Legendre basis
to enforce functional continuity at element boundaries.  If
$\phi^{(L)}=\sum_{in}C^{(L)}\chi^{(L)}_{i}(x)T_{n}(t)$ is defined in the
region from $x-\Delta x$ to $x$ and
$\phi^{(R)}=\sum_{in}C^{(R)}\chi^{(R)}_{i}(x)T_{n}(t)$ is defined in the
region from $x$ to $x+\Delta x$, functional continuity is maintained by
requiring that
\begin{equation}
\sum_{i}C^{(L)}_{in}=\sum_{i}(-1)^{i}C^{(R)}_{in} \forall n.
\label{eq:eltcontinuity}
\end{equation}

Boundary and initial conditions are specified in a similar way.  Requiring
$\phi(x,t)$ to equal $f(x,t)$ at the initial time in some element $\alpha$
yields
\begin{equation}
\sum_{n}C^{(\alpha)}_{in}(-1)^{n}=\sum_{n}f^{(\alpha)}_{in}(-1)^{n} \forall i,\alpha.
\end{equation}
Matching a function on the left boundary of some element $\alpha'$ yields
\begin{equation}
\sum_{i}C^{(\alpha')}_{in}(-1)^{i}=\sum_{i}f^{(\alpha')}_{in}(-1)^{i} \forall n,
\end{equation}
while matching the function on the right boundary of element $\alpha''$ yields
\begin{equation}
\sum_{i}C^{(\alpha'')}_{in}=\sum_{i}f^{(\alpha'')}_{in} \forall n.
\end{equation}
Ensuring functional continuity between element $\alpha'$ on the left and
$\alpha''$ on the right yields
\begin{equation}
\sum_{i}C^{(\alpha')}_{in}=\sum_{i}(-1)^{i}C^{(\alpha'')}_{in} \forall n.
\end{equation}


When projecting a function $f(x,t)$ into this basis, a projection which
enforces continuity can be found by solving the linear system 
\begin{eqnarray}
C^{(L)}_{in} + \sum_{n}\lambda_{n} &= f^{(L)}_{in} \forall i,n \\
C^{(R)}_{in} + \sum_{n}(-1)^{i}\lambda_{n} &= f^{(R)}_{in} \forall i,n \\
\sum_{i} C^{(L)}_{in}-\sum_{i}(-1)^{i}C^{(R)}_{in}&=0 \forall n.
\label{eq:continuityprojection}
\end{eqnarray}
where $f^{(\alpha)}_{in}=\int dx \int dt f(x,t)
\chi^{*(\alpha)}_{i}(x)T^{*}_{n}(t)$ is the overlap of the basis and the
expanded function over element $\alpha$.

The least action relaxation step can be performed in one of two ways.
To solve for the relaxed function directly, it is
necessary to solve the least action equations with boundary conditions given
by the trial wavefunction on the boundary of the relaxation volume.  If
$\phi(x,t)=\sum_{i,n}C_{in}\chi_{i}(\vec{x})T_{n}(t)$ is the trial
wavefunction, the relaxed wavefunction
$\phi'(x,t)=\sum_{i,n}C'_{in}\chi_{i}(\vec{x})T_{n}(t)$ is found by solving
the linear system
\begin{equation}
C^{'(\alpha)}_{in}[i O^{(\alpha \beta)}_{ij}Q_{nm} - 2 \kappa T^{\alpha
  \beta}_{ij}U_{nm}-\nu O^{\alpha \beta}_{ij}U_{nm}]=0 \forall j,m,\beta
\label{eq:leastaction_relaxedfunction}
\end{equation}
with initial condition
\begin{equation}
\sum_{n}C^{'(\alpha)}_{in}(-1)^{n}=\sum_{n}C^{(\alpha)}_{in}(-1)^{n}
\forall i,\alpha. 
\end{equation}
boundary conditions
\begin{equation}
\sum_{i}C^{'(L)}_{in}(-1)^{i}=\sum_{i}C^{(L)}_{in}(-1)^{i} \forall n,
\end{equation}
and
\begin{equation}
\sum_{i}C^{'(R)}_{in}=\sum_{i}C^{(R)}_{in} \forall n.
\end{equation}
and internal continuity enforced by
\begin{equation}
\sum_{i}C^{'(L)}_{in}=\sum_{i}(-1)^{i}C^{'(R)}_{in} \forall n.
\end{equation}
Alternatively, the least action linear system can be cast in the form of a
correction $\delta \phi(x,t)=\sum_{i,n}\delta C_{in}\chi_{i}(\vec{x})T_{n}(t)$ to a trial function $\phi(x,t)$.  If
\begin{equation}
r^{(\beta)}_{ij}=-C^{(\alpha)}_{in}[i O^{(\alpha \beta)}_{ij}Q_{nm} - 2 \kappa
T^{\alpha \beta}_{ij}U_{nm}-\nu O^{\alpha \beta}_{ij}U_{nm}] \forall j,m,\beta
\end{equation}
is the residual of the least action equations, $\delta C^{(\alpha)}_{in}$ can
be found by solving
\begin{equation}
\delta C^{(\alpha)}_{in}[i O^{(\alpha \beta)}_{ij}Q_{nm} - 2 \kappa T^{\alpha
  \beta}_{ij}U_{nm}-\nu O^{\alpha \beta}_{ij}U_{nm}]=r^{(\beta)}_{jm} \forall
j,m,\beta
\label{eq:leastaction_correction}
\end{equation}
with initial condition
\begin{equation}
\sum_{n}\delta C^{(\alpha)}_{in}(-1)^{n}=0
\forall i,\alpha, 
\label{eq:ic_correction}
\end{equation}
boundary conditions
\begin{equation}
\sum_{i}\delta C^{(L)}_{in}(-1)^{i}=0 \forall n,
\label{eq:bc1_correction}
\end{equation}
and
\begin{equation}
\sum_{i}\delta C^{(R)}_{in}=0 \forall n,
\label{eq:bc2_correction}
\end{equation}
and internal continuity enforced by
\begin{equation}
\sum_{i}\delta C^{(L)}_{in}=\sum_{i}(-1)^{i}\delta C^{(R)}_{in} \forall n.
\end{equation}

Restating the problem in the form of solving  for a correction to a trial
wavefunction has the additional advantage that all terms in the least action
linear system can be acted upon by intergrid transfer operators, making the
problem susceptible to multigrid approaches.

\section{Convergence}
The relaxation procedure described in this paper seeks to solve for the
minimum action wavefunction for a given initial condition by iteratively
decreasing the action accumulated by a series of trial wavefunctions.
Convergence to the true minimum action wavefunction is achieved when the
action can no longer be lowered, and the residual of the least action equation
is zero.

\begin{figure}
\begin{center}
\leavevmode
\includegraphics[width=0.5\textwidth]{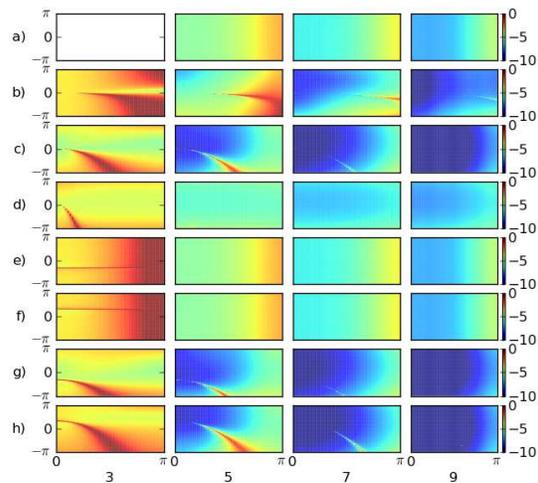}
\end{center}
\caption{(Color online) Log ratio of the residual norm before and after
  relaxation vs wavenumber (horizontal) and frequency (vertical) for trial
  wavefunction $f(x,t)=e^{i(kx+\omega t)}$, calculated using a single volume
  element.  Columns show convergence for a given number of spatial and
  temporal basis functions, while rows show convergence for different values
  of the parameters $\kappa$ and $\nu$ in
  Eq. \ref{eq:leastaction_dimensionless}.  a) $\kappa=0$, $\nu=0$, b)
  $\kappa=0.1$, $\nu=0$, c) $\kappa=1.0$, $\nu=0$, d) $\kappa=10.0$, $\nu=0$,
  e) $\kappa=0$, $\nu=1.0$, f) $\kappa=0$, $\nu=-1.0$, g) $\kappa=1.0$,
  $\nu=1.0$, h) $\kappa=1.0$, $\nu=-1.0$.  Red areas correspond to ratios
  greater than 1.0, where the relaxation procedure does not converge.}
\label{fig:residnorm_oneelt}
\end{figure}

\begin{figure}
\begin{center}
\leavevmode
\includegraphics[width=0.5\textwidth]{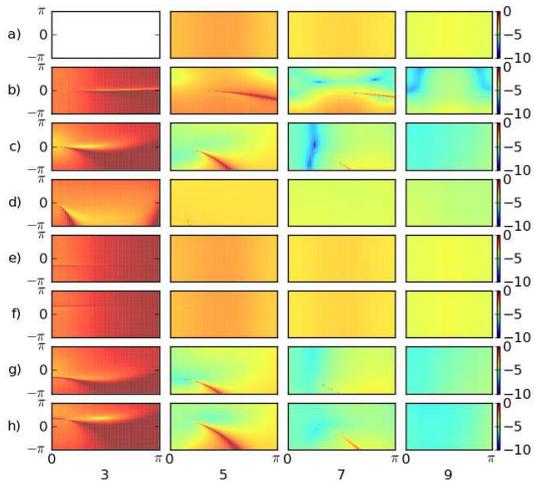}
\end{center}
\caption{(Color online) Log ratio of the magnitude of the accumulated action
  before and after relaxation vs wavenumber (horizontal) and frequency
  (vertical) for trial wavefunction $f(x,t)=e^{i(kx+\omega t)}$, calculated
  using a single volume element.  Columns show convergence for a given number
  of spatial and temporal basis functions, while rows show convergence for
  different values of the parameters $\kappa$ and $\nu$ in
  Eq. \ref{eq:leastaction_dimensionless}.  a) $\kappa=0$, $\nu=0$, b)
  $\kappa=0.1$, $\nu=0$, c) $\kappa=1.0$, $\nu=0$, d) $\kappa=10.0$, $\nu=0$,
  e) $\kappa=0$, $\nu=1.0$, f) $\kappa=0$, $\nu=-1.0$, g) $\kappa=1.0$,
  $\nu=1.0$, h) $\kappa=1.0$, $\nu=-1.0$.  Red areas correspond to ratios
  greater than 1.0, where the relaxation procedure does not converge.}
\label{fig:action_oneelt}
\end{figure}

\begin{figure}
\begin{center}
\leavevmode
\includegraphics[width=0.5\textwidth]{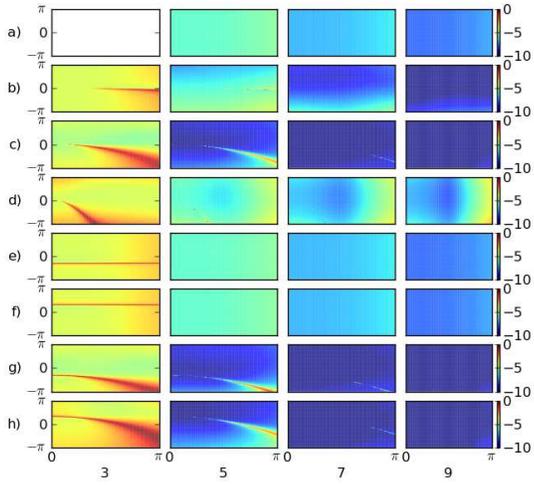}
\end{center}
\caption{(Color online) Log ratio of the residual norm before and after relaxation
  vs wavenumber (horizontal) and frequency (vertical) for trial wavefunction
  $f(x,t)=e^{i(kx+\omega t)}$, calculated using two adjacent volume elements.
  Columns show convergence for a given number of spatial and temporal basis
  functions, while rows show convergence for different values of the
  parameters $\kappa$ and $\nu$ in Eq. \ref{eq:leastaction_dimensionless}.  a)
  $\kappa=0$, $\nu=0$, b) $\kappa=0.1$, $\nu=0$, c) $\kappa=1.0$, $\nu=0$, d)
  $\kappa=10.0$, $\nu=0$, e) $\kappa=0$, $\nu=1.0$, f) $\kappa=0$, $\nu=-1.0$,
  g) $\kappa=1.0$, $\nu=1.0$, h) $\kappa=1.0$, $\nu=-1.0$.  Red areas
  correspond to ratios greater than 1.0, where the relaxation procedure does
  not converge.}
\label{fig:residnorm_twoelts}
\end{figure}

\begin{figure}
\begin{center}
\leavevmode
\includegraphics[width=0.5\textwidth]{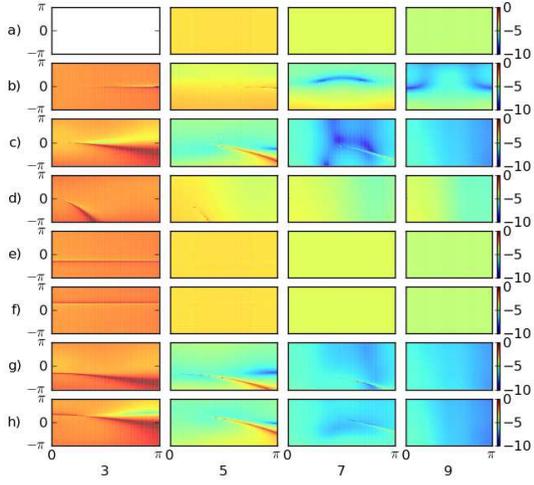}
\end{center}
\caption{(Color online) Log ratio of the magnitude of the accumulated action
  before and after relaxation vs wavenumber (horizontal) and frequency
  (vertical) for trial wavefunction $f(x,t)=e^{i(kx+\omega t)}$, calculated
  using a two adjacent volume elements.  Columns show convergence for a given
  number of spatial and temporal basis functions, while rows show convergence
  for different values of the parameters $\kappa$ and $\nu$ in
  Eq. \ref{eq:leastaction_dimensionless}.  a) $\kappa=0$, $\nu=0$, b)
  $\kappa=0.1$, $\nu=0$, c) $\kappa=1.0$, $\nu=0$, d) $\kappa=10.0$, $\nu=0$,
  e) $\kappa=0$, $\nu=1.0$, f) $\kappa=0$, $\nu=-1.0$, g) $\kappa=1.0$,
  $\nu=1.0$, h) $\kappa=1.0$, $\nu=-1.0$.  Red areas correspond to ratios
  greater than 1.0, where the relaxation procedure does not converge.}
\label{fig:action_twoelts}
\end{figure}

The convergence of the relaxed wavefunction to the minimum action solution was
tested by expanding a trial wavefunction of the form
$f(x,t)=e^{i(kx+\omega t)}$ into a Legendre basis consisting of either one
element ranging from $[-1,1]$ or two elements ranging from $[-1,0]$ and
$[0,1]$.  Convergence was measured by taking the ratio of the magnitude of the
accumulated action before and after relaxation, or alternatively by taking the
ratio of the norm of the action residuals.  Rates of convergence using one
element are shown in Figures \ref{fig:residnorm_oneelt} and
\ref{fig:action_oneelt}, while rates of convergence using two elements are
shown in Figures \ref{fig:residnorm_twoelts} and \ref{fig:action_twoelts}.
Because $\kappa$ and $\nu$ may vary widely from problem to problem, or within
the same problem due to intergrid transfer, calculations were made for a wide
range of both parameters, including $\kappa >>1$, corresponding to large
amounts of action due to high kinetic energy, and $\nu<0$, where the energy
eigenvalue spectrum spanned zero.  For all values of $\kappa$ and $\nu$
tested, the relaxation procedure yielded rapid rates of convergence in the
limit of sufficiently large basis sets.  Rates of convergence were somewhat
higher for calculations made with larger basis sets, although this must be
balanced against the higher cost of an individual relaxation step.  More
important was the requirement that the basis set yield sufficient free
parameters to allow for convergence while satisfying $N_{x}+N_{t}(2+N_{e}-1)$
constraint equations, where $N_{e}$ is the number of elements.  For
$N_{x}=N_{t}=3$ the 9 free parameters are matched by 9 constraint equations
for a single element calculation while a two element calculation yields 18
free parameters and 12 constraints.  The large number of constraint equations,
with corresponding action costs in the form of Lagrange multipliers, means
that rates of convergence may be slow or even diverging for small basis sets.
Higher order basis sets yielded rapid convergence over nearly all of parameter
space, with convergence slowing only in the vicinity of the
$-w=\frac{1}{2}\kappa^{2}+\nu$ curve where the trial solution is already a
good approximation to the minimum action wavefunction.

Comparing these figures, it is apparent that relaxing over two elements
simultaneously is very inefficient.  As the cost of solving a linear system
for n unknowns scales as $n^{3}$, relaxing over 2 volumes simultaneously is
approximately 4 times as expensive as relaxing over each volume separately.
However, comparing figures \ref{fig:residnorm_oneelt} and
\ref{fig:residnorm_twoelts}, or \ref{fig:action_oneelt} and
\ref{fig:action_twoelts} shows that this added expense does not
result in appreciably faster rates of convergence.  Despite this inefficiency,
the inability of the relaxation procedure to change the wavefunction on the
boundary of the relaxation volume means that a convergent procedure must at
some point relax over volumes larger than a single element.

\begin{figure}
\begin{center}
\leavevmode
\includegraphics[width=0.5\textwidth]{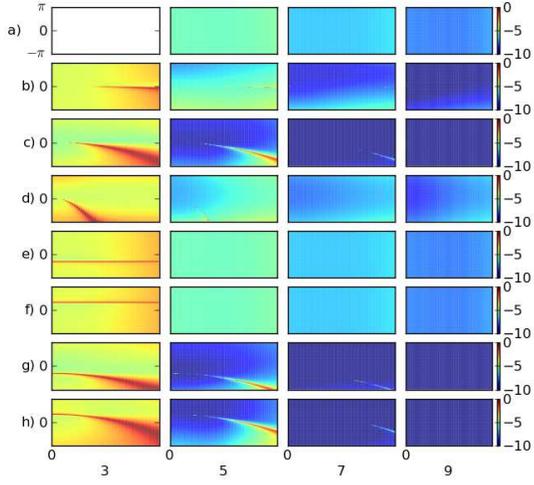}
\end{center}
\caption{(Color online) Log ratio of the residual norm before and after
  relaxation vs wavenumber (horizontal) and frequency (vertical) for trial
  wavefunction $f(x,t)=e^{i(kx+\omega t)}$, calculated for two adjacent
  elements using the intergrid transfer method.  Columns show convergence for
  a given number of spatial and temporal basis functions, while rows show
  convergence for different values of the parameters $\kappa$ and $\nu$ in
  Eq. \ref{eq:leastaction_dimensionless}.  a) $\kappa=0$, $\nu=0$, b)
  $\kappa=0.1$, $\nu=0$, c) $\kappa=1.0$, $\nu=0$, d) $\kappa=10.0$, $\nu=0$,
  e) $\kappa=0$, $\nu=1.0$, f) $\kappa=0$, $\nu=-1.0$, g) $\kappa=1.0$,
  $\nu=1.0$, h) $\kappa=1.0$, $\nu=-1.0$.  Red areas correspond to ratios
  greater than 1.0, where the relaxation procedure does not converge.}
\label{fig:residnorm_twoelts_smooth_rough}
\end{figure}

\begin{figure}
\begin{center}
\leavevmode
\includegraphics[width=0.5\textwidth]{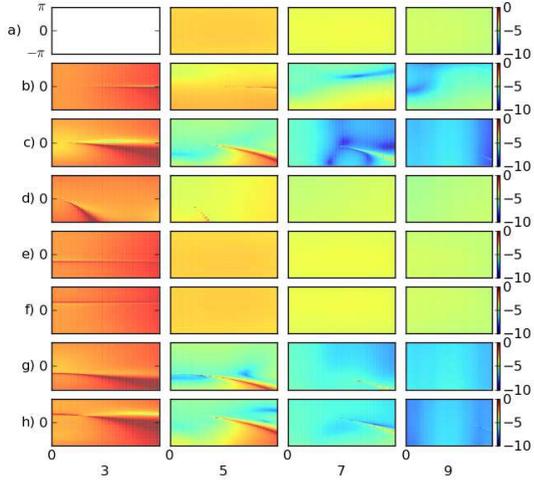}
\end{center}
\caption{(Color online) Log ratio of the magnitude of the accumulated action
  before and after relaxation vs wavenumber (horizontal) and frequency
  (vertical) for trial wavefunction $f(x,t)=e^{i(kx+\omega t)}$, calculated
  for two adjacent elements using the intergrid transfer method.  Columns show
  convergence for a given number of spatial and temporal basis functions,
  while rows show convergence for different values of the parameters $\kappa$
  and $\nu$ in Eq. \ref{eq:leastaction_dimensionless}.  a) $\kappa=0$,
  $\nu=0$, b) $\kappa=0.1$, $\nu=0$, c) $\kappa=1.0$, $\nu=0$, d)
  $\kappa=10.0$, $\nu=0$, e) $\kappa=0$, $\nu=1.0$, f) $\kappa=0$, $\nu=-1.0$,
  g) $\kappa=1.0$, $\nu=1.0$, h) $\kappa=1.0$, $\nu=-1.0$.  Red areas
  correspond to ratios greater than 1.0, where the relaxation procedure does
  not converge.}
\label{fig:action_twoelts_smooth_rough}
\end{figure}

A more efficient way of relaxing over large volumes makes use of
intergrid transfer operators.  In Figures
\ref{fig:residnorm_twoelts_smooth_rough} and
\ref{fig:action_twoelts_smooth_rough}, relaxation over two adjacent volumes
was accomplished by first transferring the residual from two fine elements to
one coarse element using Eq. \ref{eq:residual_rough}.  The correction found on
the coarse grid was then transferred back to the fine grid using
Eq. \ref{eq:correction_fine}.  This partially relaxed wavefunction was then
relaxed over the two fine elements separately.  In all, this process involved
solving 3 linear systems of size n, so that the computational effort was 1.5
times that of relaxing over each volume separately, rather than 4 times as in
the case of simultaneous relaxation.  Despite this decrease in
computational effort, rates of convergence were comparable to those obtained
by relaxing over both volumes at once.  The decrease in computational
effort is more pronounced in higher dimensions: using intergrid transfer to
relax over a hypercube of $2^{d}$ elements increases costs by a factor of
$1+2^{-d}$ relative to relaxing over each element separately, while relaxing
over the entire hypercube simultaneously increases costs by a factor of
$2^{2d}$.

\section{Conclusions}
Efficiently and accurately propagating a time dependent wavefunction is a
fundamental problem of computational quantum mechanics, with applications
ranging across many areas of physics.  This paper has addressed this problem
by developing an iterative procedure for relaxing a trial wavefunction toward
a variationally optimum, action minimizing solution.  This relaxation
procedure is trivially parallelizeable for problems involving a local
Hamiltonian, and does not rely on the Hamiltonian being positive definite,
making it well suited for incorporation into multigrid relaxation methods.  A
local Fourier analysis shows that this procedure yields rapid rates of
convergence over a wide range of parameters in the limit that the basis set
defined over a particular volume element is sufficiently large.

Although the discussion in this paper has focused on the specific problem of
the time dependent Scr\"odinger equation, the analysis depends primarily on
the existence of a local Hamiltonian.  As this is a very common condition in
physical problems, it may in the future be desirable to extend this analysis
to other computationally difficult problems.

\end{document}